\documentclass[aps,superscriptaddress,twoside,twocolumn,floatfix,a4paper,longbibliography]{revtex4-1}

\usepackage{amsmath}
\usepackage{amssymb}
\usepackage[utf8]{inputenc}
\usepackage[T1]{fontenc}
\usepackage[separate-uncertainty=true]{siunitx}
\usepackage[export]{adjustbox}
\usepackage{float}
\usepackage{color}
\usepackage[ocgcolorlinks,citecolor=red]{hyperref}

\begin{document}

\newcommand{\orcid}[1]{\href{https://orcid.org/#1}{\includegraphics[width=8pt]{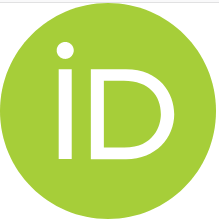}}}

\author{E. Strambini\orcid{0000-0003-1135-2004}}
\email{elia.strambini@sns.it}
\affiliation{NEST, Istituto Nanoscienze-CNR and Scuola Normale Superiore, I-56127 Pisa, Italy}

\author{M. Spies\orcid{0000-0002-3570-3422}}
\email{maria.spies@nano.cnr.it}
\affiliation{NEST, Istituto Nanoscienze-CNR and Scuola Normale Superiore, I-56127 Pisa, Italy}

\author{N. Ligato\orcid{0000-0001-7655-9976}}
\affiliation{NEST, Istituto Nanoscienze-CNR and Scuola Normale Superiore, I-56127 Pisa, Italy}

\author{S. Ilic\orcid{0000-0003-1406-9407}}
\affiliation{Centro de F\'isica de Materiales (CFM-MPC) Centro Mixto CSIC-UPV/EHU, E-20018 Donostia-San Sebasti\'an,  Spain}

\author{M. Rouco\orcid{0000-0003-2175-9238}}
\affiliation{Centro de F\'isica de Materiales (CFM-MPC) Centro Mixto CSIC-UPV/EHU, E-20018 Donostia-San Sebasti\'an,  Spain}

\author{Carmen González Orellana\orcid{0000-0003-4033-5932}}
\affiliation{Centro de F\'isica de Materiales (CFM-MPC) Centro Mixto CSIC-UPV/EHU, E-20018 Donostia-San Sebasti\'an,  Spain}

\author{Maxim Ilyn\orcid{0000-0001-5832-2449}}
%\email{maxim.ilin@ehu.eus}
\affiliation{Centro de F\'isica de Materiales (CFM-MPC) Centro Mixto CSIC-UPV/EHU, E-20018 Donostia-San Sebasti\'an,  Spain}

\author{Celia Rogero\orcid{0000-0002-2812-8853}}
\affiliation{Centro de F\'isica de Materiales (CFM-MPC) Centro Mixto CSIC-UPV/EHU, E-20018 Donostia-San Sebasti\'an,  Spain}
\affiliation{Donostia International Physics Center (DIPC), 20018 Donostia--San Sebasti\'an, Spain}

\author{F.S. Bergeret\orcid{0000-0001-6007-4878}}
\affiliation{Centro de F\'isica de Materiales (CFM-MPC) Centro Mixto CSIC-UPV/EHU, E-20018 Donostia-San Sebasti\'an,  Spain}
\affiliation{Donostia International Physics Center (DIPC), 20018 Donostia--San Sebasti\'an, Spain}
\affiliation{Institute of Solid State Theory, Wilhelm-Klemm-Straße 10, University of Münster, 48149 Münster, Germany}
%\email{fs.bergeret@csic.es}

\author{J. S. Moodera\orcid{0000-0002-2480-1211}}
\affiliation{Physics Department and Plasma Science and Fusion Center, Massachusetts Institute of Technology, Cambridge, Massachusetts 02139, USA}

\author{P. Virtanen\orcid{0000-0001-9957-1257}}
\affiliation{Department of Physics and Nanoscience Center, University of Jyväskylä, P.O. Box 35 (YFL), FI-40014 University of Jyväskylä, Finland}

\author{T. T. Heikkilä\orcid{0000-0002-7732-691X}}
\affiliation{Department of Physics and Nanoscience Center, University of Jyväskylä, P.O. Box 35 (YFL), FI-40014 University of Jyväskylä, Finland}

\author{F. Giazotto\orcid{0000-0002-1571-137X}}
\email{francesco.giazotto@sns.it}
\affiliation{NEST, Istituto Nanoscienze-CNR and Scuola Normale Superiore, I-56127 Pisa, Italy}

%\title{Rectification in a Eu-chalcogenide-based superconducting diode}
\title{Superconducting spintronic tunnel diode}

\maketitle

\textbf{Diodes are key elements for electronics, optics, and detection. 
The search for a material combination providing the best performances for the required application is continuously ongoing. 
Here, we present a superconducting spintronic tunnel diode based on the strong spin filtering and splitting generated by an EuS thin film between a superconducting Al and a normal metal Cu layer. The Cu/EuS/Al tunnel junction achieves a large rectification (up to $\sim40$\%) already for a small voltage bias ($\sim 200$ $\mu$V) thanks to the small energy scale of the system: the Al superconducting gap. With the help of an analytical theoretical model we can link the maximum rectification to the spin polarization ($P$) of the barrier and describe the quasi-ideal Schottky-diode behavior of the junction.
This cryogenic spintronic rectifier is promising for the application in highly-sensitive radiation detection for which two different configurations are evaluated. In addition, the superconducting diode may pave the way for future low-dissipation and fast superconducting electronics.
} 
%-------------INTRO, max 500 words---------
Diodes are non-linear and non-reciprocal circuits in which a lack of spatial inversion symmetry provides a strongly direction-selective electron transport. In the long and successful history of diodes, the material search for this symmetry breaking has been mainly focused on semiconducting and metallic junctions. 
However, owing to their large energy gap, semiconductors cease to work at the sub-Kelvin temperatures relevant for emerging cryogenic electronics \cite{braginski2019superconductor} and ultrasensitive detection, especially at sub-THz frequencies~\cite{farrah_review_2019}. 
This problem could be partially solved by using low-dimensional structures like quantum dots, which do exhibit current rectification \cite{ono2002current,dicarlo2003photocurrent}. Given that the electron-hole symmetry in quantum dots can be tuned only within the level of a single quantum channel, the impedance of such systems tends to be high, and the rectified currents thereby very low, limiting the value of this approach.
Superconductors would be ideal candidates for the realization of cryogenic diodes and detectors due to their intrinsic low impedance, and the lower energy scales of the superconducting gap ($\sim$~meV) compared to semiconductors ($\sim$~eV). 
Still, the implementation of a superconducting diode turns out to be difficult since it requires breaking of the electron-hole symmetry, whereas the BCS superconducting state is, by construction, electron-hole symmetric.
Recently, supercurrent diodes have been engineered with metallic superlattices in strong magnetic fields offering the required lack of spatial inversion \cite{ando_observation_2020} or with unconventional Josephson junctions  \cite{baumgartner_josephson_2021,wu_realization_2021}. 

Here, we show an alternative approach of realizing a superconducting diode based on a spin-selective EuS tunnel junction.  Eu-based chalcogenides combined with superconductors offer bright perspectives for the realization of novel technologies based on the interplay between superconductivity and ferromagnetism. This can be realized in thin film bilayers which consist of ferromagnetic insulator (FI) and superconductor (S) materials. They can show ideal spin filtering and spin splitting \cite{meservey_spin-polarized_1994,giazotto_superconductors_2008}, as already demonstrated in a number of seminal  experiments performed on FI/S-based tunnel junctions~\cite{moodera_electron-spin_1988,meservey_spin-polarized_1994,strambini_revealing_2017}.
Recently, it has been shown that when both spin filtering and splitting are present in FI/S tunnel junctions, it is possible to break the electron-hole
symmetry of the system and generate direction-selective electron transport \cite{giazotto2015ferromagnetic}, which is at the basis of charge rectification and thermoelectricity \cite{ozaeta_predicted_2014,machon_nonlocal_2013,giazotto_very_2015}. 
That makes the design of the present superconducting spintronic device a promising approach for the implementation of biasless ultrasensitive THz detectors \cite{heikkila_thermoelectric_2018}.

%----------------------Fig1-------
\begin{figure*}[ht!]
\includegraphics[center,scale=0.50]{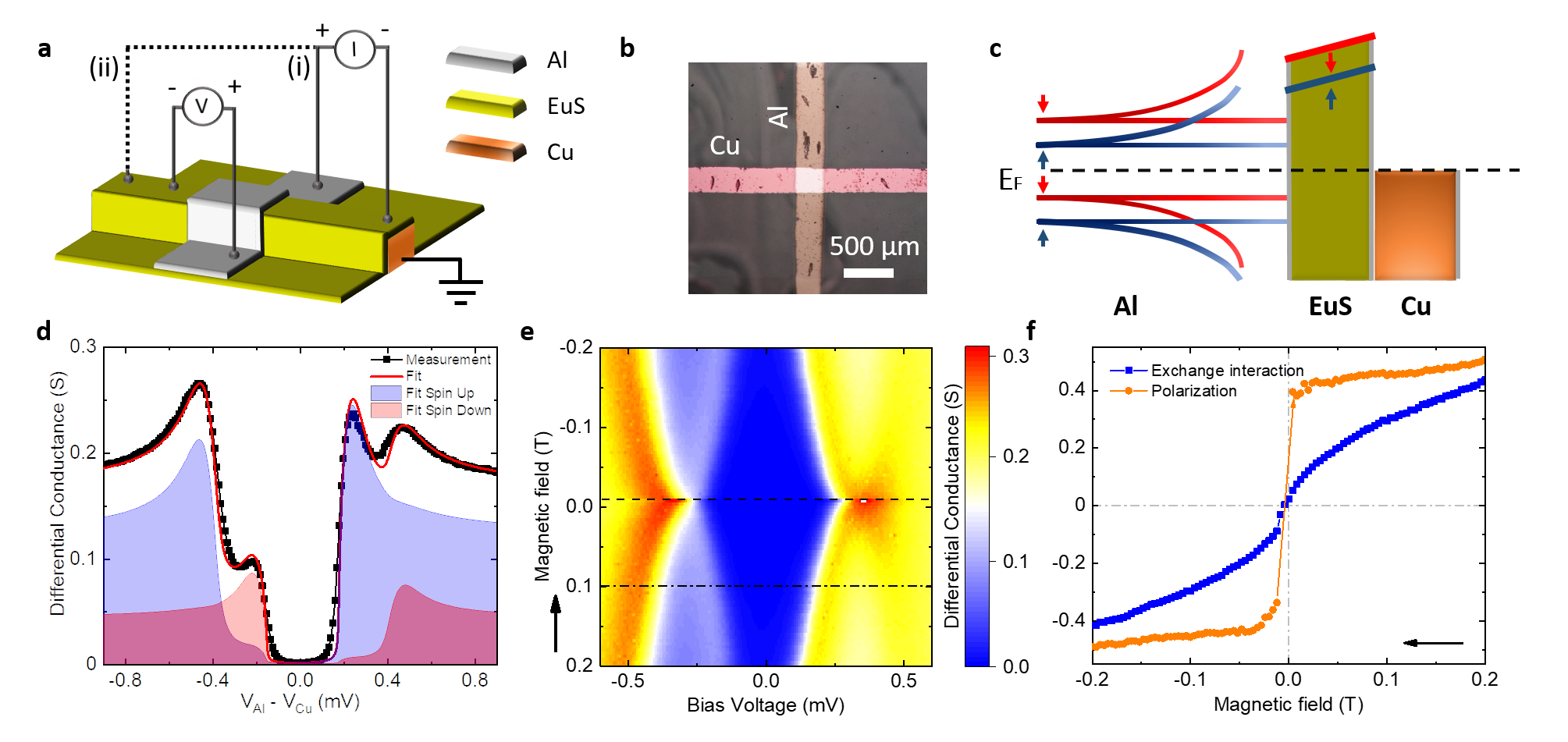}
\caption{
\textbf{Working principle and characteristics of the superconducting diode.}
\textbf{a}, Schematic of the device structure: a Cu strip (orange) is covered by a EuS layer (green) and a perpendicular Al strip (gray). Measurement setups: The electric current is applied (i) from the Al to the Cu strip or (ii) along the Cu strip. The voltage drop is measured between the Al and the Cu strip on the remaining two wires of the four-wire set-up. 
\textbf{b}, Visible light microscopy image of the device. 
\textbf{c}, Schematic of the DoS along the vertical axis of the structure (Al/EuS/Cu from top to bottom). The dashed line indicates the Fermi level. Note that the EuS layer induces spin splitting in the superconducting DoS, and spin filtering thanks to the different height of the tunnel barrier for the two spin species. The red (blue) line corresponds to spin up (down) DoS in the Al layer.
\textbf{d}, Exemplary differential conductance (black) measured as a function of voltage across the junction at an applied external magnetic field $B$ of $\SI{0.1}{\tesla}$ at $\simeq \SI{100}{\milli\kelvin}$. By employing a numerical model (detailed in the Methods section, Eq.~\ref{eq:diode-eq-I} and Eq.~\ref{eq:DOS-0z}), the fit for the differential conductance (red) and the contributions of the spin up (light blue) and spin down (light red) electrons were calculated with these fitting parameters: $\Delta_0 = 0.33$meV, $h=0.32 \Delta_0$, $P=0.48$, $\Gamma=0.01 \Delta_0$, $T=300 $mK. \textbf{e}, Color map of the differential conductance $dI/dV(V)$ measured for $B$ ranging from $\SI{-0.2}{\tesla}$ to $\SI{0.2}{\tesla}$. The sweep direction is indicated by the arrow. The data in panel \textbf{d} corresponds to the dash-dotted line (B= 0.1 T). The coercive field at the temperature of this measurement (100 mK) corresponds to -9 mT, indicated by a dashed line.
\textbf{f}, Exchange field ($h$) induced in the superconducting Al strip (blue) and polarization ($P$) of the EuS tunnel barrier as a function of the external magnetic field $B$. Both quantities are extracted from the best fitting results of the data as shown in panel \textbf{d}. The sweep direction is again indicated by an arrow.} 
\label{fig1}
\end{figure*} 
%----------------FIG2-------------------------
\begin{figure*}[ht!]
\includegraphics[center, scale=0.6]{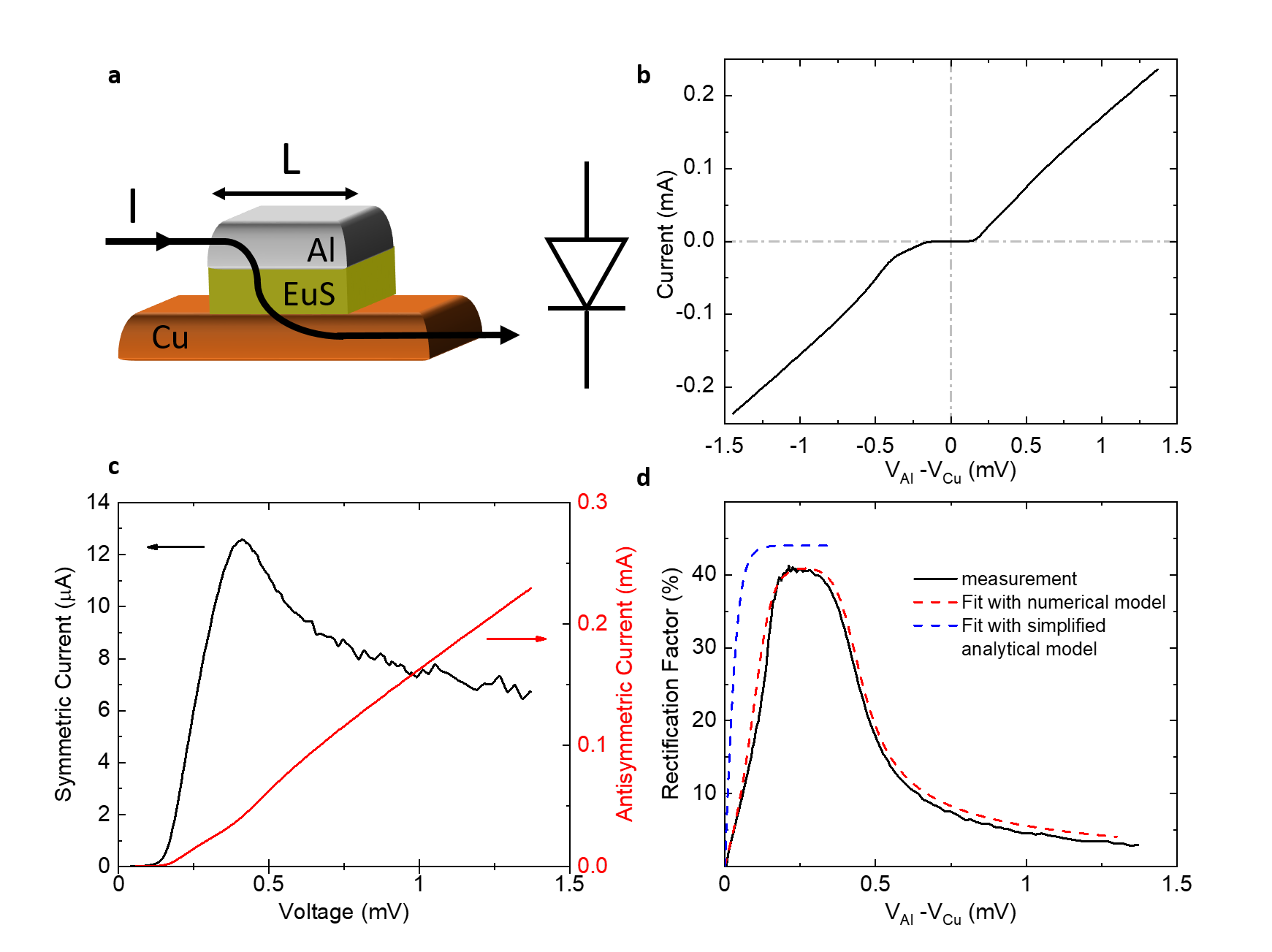}
\caption{\textbf{Rectification of the superconducting diode.}
\textbf{a}, Schematic of the N/FI/S tunnel junction. The path of the tunneling current is indicated by the black line and its arrows. In terms of electronic circuit elements this junction behaves like the indicated diode: the current flows preferentially from the Al layer to the Cu layer while the reverse flow is inhibited. 
\textbf{b}, Current-to-Voltage ($I(V)$) characteristics of the junction measured at $T\simeq \SI{50}{\milli\kelvin}$, $B= \SI{0.1}{\tesla}$ in the four-wire configuration (i).
\textbf{c}, Symmetric and antisymmetric parts of the $I(V)$ characteristic of panel \textbf{c} showing a sizable symmetric component of the current.
\textbf{d}, Rectification coefficient $R(V)=I_{\rm Sym}(V)/I_{\rm Antisym}(V)$ evaluated from \textbf{e} (black line) along with the comparison with the rectification extracted from the approximated analytical model $R=P\tanh[eV/(2K_B T)]$ (blue line)  and the full numerical ones (red line). Notice the good qualitative agreement with the simplified model predicting the saturation at $R\simeq P \sim 40\%$ at 225 - 280 $\mu$V. The model ceases to work when $eV \gtrsim \Delta-h \sim 250$ $\mu$eV. The discrepancy between the analytical model and the experiment mostly comes from weak inelastic scattering, and to a lesser extent from spin relaxation and orbital depairing, as shown by the full numerical calculations.
}
\label{fig2}
\end{figure*}

The working principle and device characteristics of the normal metal–ferromagnetic insulator–superconductor (N/FI/S) tunnel junction, central to this paper, are shown in Fig.~\ref{fig1}. The schematic of the device structure and measurement configurations for the tunnel spectroscopy can be seen in panel (a). A normal metal strip of Cu and a S strip of Al are oriented perpendicular to one another forming a cross-bar, and are separated by a FI barrier of EuS (see Methods for fabrication details). The EuS layer induces a spin splitting by an energy  with magnitude ($h$) in S through interface exchange interaction \cite{diepen_nuclear_1968,moodera_electron-spin_1988,hijano_coexistence_2021}, and its FI nature causes a spin filtering ($P$) of the electron tunneling across the junction.
The former results in an opposite energy shift for the BCS density of states (DoS) of the two spin species, as sketched in Fig.~\ref{fig1}(c), while the latter forms a tunneling barrier with different height for the two spin species.
This twofold effect can be probed experimentally by measuring the differential conductance of the tunnel junction and leads to qualitative changes in the system's transport characteristics \cite{bergeret2018colloquium,heikkila2019thermal,moodera_electron-spin_1988}. 
An example of a tunneling conductance measurement as a function of bias voltage across the N/FI/S junction is shown in Fig.~\ref{fig1}(d). At small voltages ($|V|\lesssim 200$ $\mu$V) the conductance is strongly suppressed due to the lack of states within the superconducting energy gap. At higher bias voltage four distinct peaks can be observed in total, corresponding to the four peaks of the two BCS DoS at $e|V|=\Delta\pm h$. The different amplitudes of the conductance peaks stem from the spin filtering $P$, promoting one spin channel with respect to the other. All these parameters can be extracted by fitting the conductance with a numerical model (see Eq.~\ref{eq:IV} and \ref{eq:DOS-0z} in the "Methods" for  details on the model) that takes into account the spin splitting, spin relaxation and orbital depairing due to the magnetic field~\cite{heikkila2019thermal}, as shown by the red curve of Fig.~\ref{fig1}(d).
Additionally, the application of an external magnetic field can strengthen the polarization of the EuS layer and enhance both $h$ and $P$, as shown in Fig.~\ref{fig1}(e) and (f). Notably, thanks to the ferromagnetism of the EuS, a sizable splitting and polarization are observed even at zero field ($h_0\simeq0.025 $ $\Delta$, $P_0\simeq0.2$ and $\Delta=370  \mu$eV). These vanish at the EuS coercive field ($\simeq 10$~mT).

In the following, two measurement configurations (sketch in Fig.~\ref{fig1}a) have been adopted to quantify the diode characteristics.
In configuration (i) the current flows from the S to the N layer, thereby effectively crossing the junction. A conventional rectification is observed in this case. In configuration (ii) the current flows along the N strip, and a transverse rectification is observed. In both cases the voltage drop is measured from the S to the N layer across the tunnel junction. Measurements of the two configurations are compared and discussed. 

%----------------FIG3--------
\begin{figure*}[ht!]
\includegraphics[center, scale=0.5]{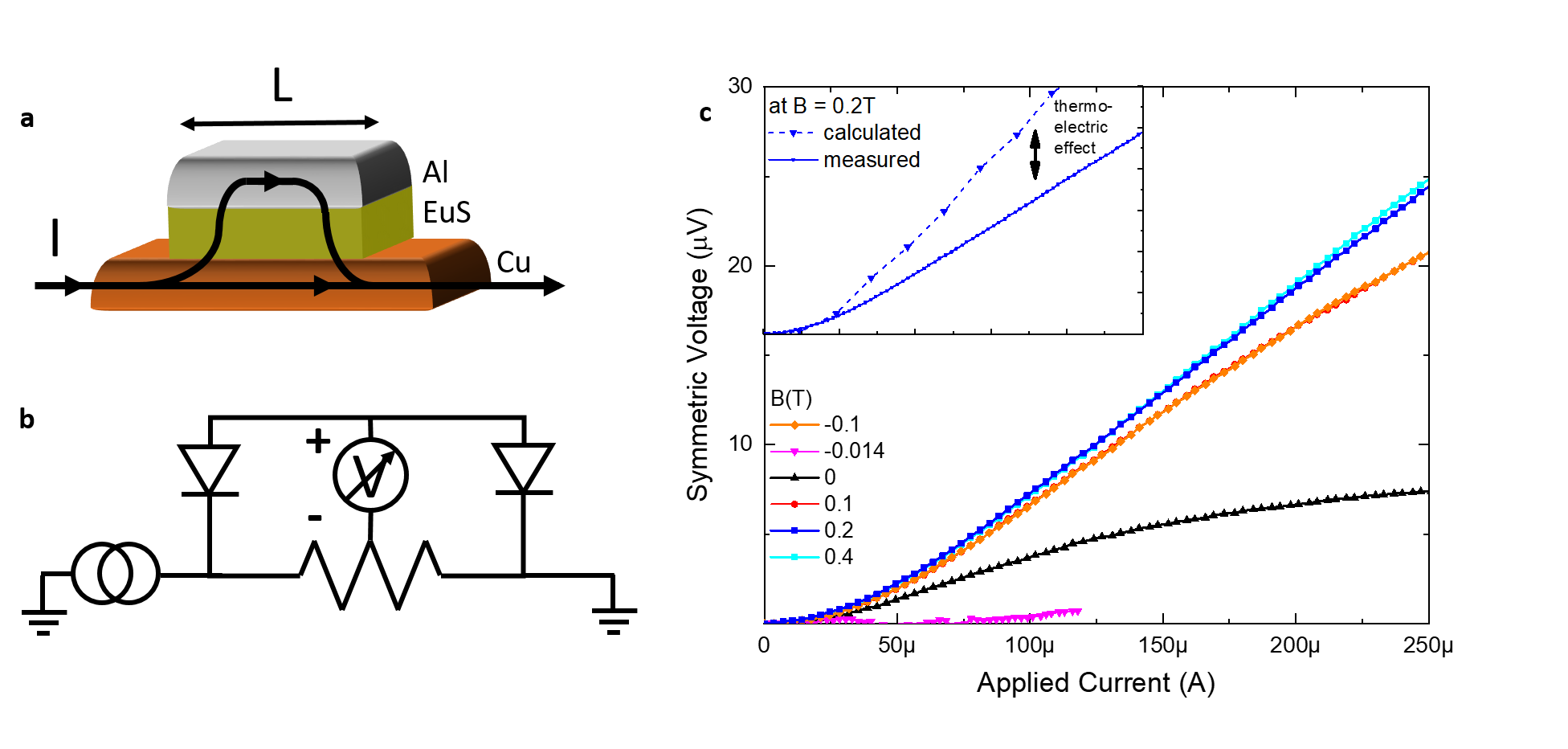}
\caption{
\textbf{Transverse rectification of the superconducting diode.
}
\textbf{a}, Schematic of the N/FI/S tunnel junction and current path. A biasing current $I_{\rm Cu}$ is applied from one end of the Cu strip to the other, while the voltage drop across the junction is measured from the Al contact to the Cu one. The path of tunneling current is indicated by the black line and its arrows. 
\textbf{b}, Electronic circuit diagram of the setup. Note that the EuS layer effectively acts as a twofold rectifier for the two incoming and outgoing currents tunneling through the FI barrier. 
\textbf{c}, Transverse voltage drop $V_{\rm sym}(I_{\rm Cu})$ measured across the barrier as a function of the applied current $I_{\rm Cu}$ at different $B$ and at $\SI{50}{\milli\kelvin}$.
Note that even at zero applied magnetic field (black curve) a voltage drop occurs, while at the coercive field ($B\simeq \SI{-14}{\milli\tesla}$) the signal is zero due to the non-polarization of the EuS layer. The $V(I)$ was symmetrized in order to discard the Ohmic (linear) component originating from the N lead. 
In the inset, the $V_{\rm sym}$ measured at $\SI{0.2}{\tesla}$ is compared with the calculated data points obtained through a theoretical model of the circuit and using the rectification value obtained from the experimental data.}
\label{fig3}
\end{figure*}

A typical current vs voltage ($I(V)$) characteristic of the tunnel junction shows a conventional rectification, as can be seen in Fig.~\ref{fig2}. It corresponds to measurement configuration (i) in voltage bias. The current bias configuration is considered in the Supplementary Information (section I) together with an alternative choice of material layers, namely EuS/Al/AlOx/Co (section II). 
The presence of the superconducting gap can be clearly recognized in the $I(V)$ characteristic displayed in Fig.~\ref{fig2}(b) with the absence of current flow at low bias, and an Ohmic behavior for relatively large voltage ($eV\gg\Delta$). 
In an intermediate voltage range, non-linearities and non-reciprocity appear, which can be visualized in the symmetric and antisymmetric parts of the $I(V)$ characteristic.
They are defined as $I_{\rm Sym} = \frac{I(V)+I(-V)}{2}$ and $I_{\rm Antisym} = \frac{I(V)-I(-V)}{2}$, and are shown in  Fig.~\ref{fig2}(c). 
The sizable $I_{\rm Sym}(V)$ already suggests an efficient charge rectification, i.e., the capability to convert an AC input into a DC output signal. 
Rectification ($R$) of a circuit can be defined as the ratio between the difference of the forward and backward current divided by the sum of the two, $R(V)= \frac{I(V)-I(-V)}{I(V)+I(-V)}= I_{\rm Sym}/I_{\rm Antisym}$, and is shown in Fig.~\ref{fig2}(d). For ideal rectifiers $R=1$, while for $R=0$ no rectification is present. In the junction a $R$ up to $\sim$40\% can be achieved in the intermediate voltage range ($eV\sim \Delta$).
This upper limit, equivalent to the polarization $P$ of the EuS junction can be understood using a simple analytical model for the N/FI/S tunnel junctions, which neglects spin-dependent scattering and orbital depairing. Within these approximations the $I(V)$ tunneling current can be simplified to the instructive expression:
\begin{equation}
    I(V) = I_S\left(e^{eV/(k_B T)}-1\right) +
    I_S\left[\cosh\left(\frac{eV}{k_B T}\right)-1\right](P-1).
    \label{eq:IV}
\end{equation}
The current scale $I_S$ depends on the physical characteristics of the device, such as the normal-state resistance, superconducting energy gap and the exchange field, as detailed in the Methods section, Eq.~\ref{eq:Is}.
The expression is valid at low temperatures ($k_B T \ll h$) and voltages ($e|V| \ll \Delta-h$).  Note that subgap states due to inelastic scattering can provide an additional contribution to the current $\delta I$, which also satisfies $\delta I(V)\neq -\delta I(-V)$, and becomes particularly important at very low temperatures (see Methods for more details). Equation \eqref{eq:IV} is composed by two elements. The first one represents the Shockley ideal diode equation \cite{shockley1949theory} and dominates when $P$ is close to unity. It describes the asymmetric I(V) curves characteristic of diodes.
The second contribution is the first correction to an ideal diode due to the non-ideal spin polarization. This yields the simple result for the rectification, $R=P\tanh[eV/(2 k_B T)]$. The maximum rectification at $|eV| \gtrsim 2 k_B T$ is hence dictated by the spin filtering efficiency $P$.
Due to the strong asymmetry induced by the spin filtering for this specific junction, $R$ is maximized around 225 - 280 $\mu$V where it obtains values as high as $\sim$40\%, in good agreement with the polarization value extracted from the $dI/dV$ fits (see Fig.~\ref{fig1}(f)). 

%----------------FIG4----------
\begin{figure*}[ht]
\includegraphics[center, scale=0.55]{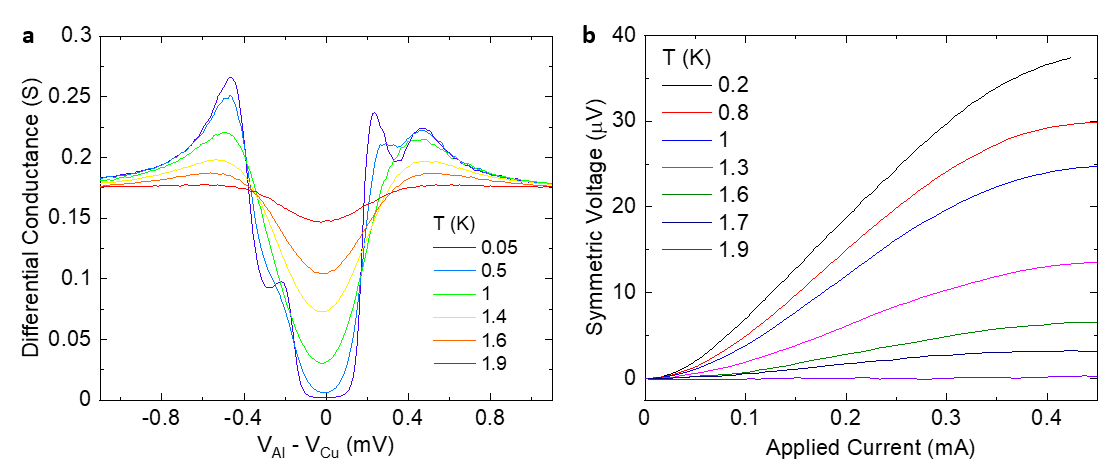}
\caption{\textbf{Temperature dependence of the superconducting rectifier.}
\textbf{a}, Differential conductance vs voltage of the junction measured for different temperatures from $\SI{50}{\milli\kelvin}$ to $\SI{1.9}{\kelvin}$.
\textbf{b}, Temperature evolution of the transverse rectification voltage vs biasing current. Both measurements are performed at $B = \SI{0.1}{\tesla}$.}
\label{fig4}
\end{figure*}

Notably, the geometry of the device together with the small resistance of the tunnel junction allow for the implementation of a "three-terminal" diode in which the path of the rectified signal (in this case the voltage) is decoupled with respect to the excitation current ($I_{\rm Cu}$) path.
This corresponds to  measurement configuration (ii)  and is sketched in Fig.~\ref{fig3}(a) and (b).
The device is operated with a current bias $I_{\rm Cu}$ applied along the Cu bottom lead, while the voltage drop is measured orthogonal to it.
At the junction, $I_{\rm Cu}$ can partially flow in the S leads and generate a voltage due to the non-symmetric response of the junction to the flowing current (see the sketch displayed in Fig.~\ref{fig3}(a)).
The resulting measured voltage $V_{\rm sym}$, symmetrized to discard the trivial ohmic component originated in the Cu lead, is shown in Fig.~\ref{fig3}(c) for different magnetic fields. 
A monotonic increase of $V_{\rm sym}(I_{\rm Cu})$ is visible and more pronounced at large fields due to the larger $h$ and $P$ of the junction. Notably, a sizable transverse rectification is present also at zero field thanks to the ferromagnetic nature of the FI layer. This characteristic is especially relevant for applications since no additional magnetic field lines need to be integrated while operating the device.
On the other hand, at the EuS coercive field ($\simeq\SI{14}{\milli\tesla}$ at base temperature) the rectified signal is not visible confirming the spintronic nature of this effect. 
From the experimental $I(V)$ characteristics of the diode shown in Fig.~\ref{fig2}(b) it is possible to model the transverse response of the diode (see Methods for calculation details). 
Our theoretical calculations compared with the data shown in the inset of Fig.~\ref{fig3}(c) are in agreement with the experiment but are generally larger than the measured data by about $\sim 30\%$. 
This difference likely stems from the thermoelectric effect that, similar to rectification, is also present in the junction with  lack of electron-hole symmetry~\cite{ozaeta_predicted_2014}. 

From a thermal model that considers the Joule heating induced by $I_{\rm Cu}$ we can estimate the resulting thermovoltage, and find that is smaller and of the opposite sign with respect to the rectification voltage (see section V of the Supplementary Information for more details), therefore confirming the co-presence of thermoelectricity in the junction.
Notably, the relative amplitude of the two effects depends on the length of the tunnel junction, with transverse rectification dominating for junctions longer than $\sim 100 \mu$m (this length scale depends on various sample specific parameters as described by Eq.~S12 in the Supplementary Information). 

Figure~\ref{fig4} shows the temperature dependence of the differential conductance and the transverse rectification voltage of the discussed tunnel junction. Notably, despite the evident thermal broadening of the $dI/dV(V)$ (see Fig.~\ref{fig4}(a)), the transverse rectification is only marginally affected below 1~K  (see Fig.~\ref{fig4}(b)), making the effect very robust even at a temperature up to nearly half of the Al critical temperature ($T_{C} \simeq 2.3$ K). However, for temperatures larger than $T_{C}/2$, a clear damping of the signal is visible with measurable effects up to $\sim$1.9~K. 
This high temperature range of operation makes our superconducting diode appealing for superconducting electronics schemes  where robustness against temperature fluctuations is desirable. Moreover, this behavior is expected to hold for other superconducting materials. There are several FI/S bilayer systems whose $T_{C}/2$ lies above 4 K (for instance, GdN/NbN bilayers~\cite{senapati_spin-filter_2011}). These materials have the advantage that they can be operated at standard $^4$He cryogenic temperatures and deposited with large-scale sputtering systems.

%----------------Discussion and Conclusion
\section*{Conclusion}
In conclusion, we have shown the capabilities of a N/FI/S tunnel junction to function both as a conventional diode and as a transverse rectifier~\cite{patent_comment}. 
The transverse rectifier benefits from a lower impedance and a direct decoupling between the AC excitation line (the antenna) and the DC sensing line. Both superconducting rectifiers can be operated in zero applied magnetic field showing promising sensitivities up to 
$\sim \SI[per-mode=symbol]{2e3}{\ampere\per\watt}$ and noise equivalent power down to 
$\sim \SI[per-mode=symbol]{1e-19}{\watt\per\sqrt{\hertz}}$ (see section III of the Supplementary Information for details on the analysis). This is a step towards the development of detectors in the THz region contributing to the terahertz gap closure.

Besides detection and rectification, this device can be used also for other conventional diode functionalities, but at much lower voltage and thereby much lower dissipation levels than conventional semiconductor-based diodes. Such applications include mixers, reverse current regulators, voltage clamping and more passive superconductive electronics~\cite{braginski2019superconductor}. Further functionalities can also be expected with more complicated structures containing several EuS or Al layers~\cite{rouco2019charge}.
%----------------Methods Section
\section*{Methods}
\textbf{Sample fabrication and transport measurements}. 
The samples are cross-bars made by electron-beam evaporation employing an in-situ shadow mask. The structures consist of a glass substrate on which the layers of Cu(20)/ EuS(4)/ Al(4)/ Al\textsubscript{2}O\textsubscript{3}(13) are deposited subsequently (thicknesses in nm). The overlap between the Al and the Cu strip has an area of $~300\times300$ $\mu$m$^2$. 
The tunneling spectroscopy is carried out at cryogenic temperatures down to 50 mK in a filtered cryogen-free dilution refrigerator. The $I(V)$ characteristics are obtained from DC four-wire measurements, as sketched in Fig.~\ref{fig1}(a), and are used to calculate the differential conductance via numerical differentiation.
\newline
\newline
\textbf{Diode equation}.
The $I(V)$ characteristic of the spin-polarized junction to a spin-split superconductor is given by (here, $e=k_B=\hbar=1$ for brevity)
\begin{equation*}
    I(V) = \sum_{\sigma} G_\sigma \int d\epsilon N_\sigma(\epsilon)[f_0(\epsilon-V)-f_0(\epsilon)],
    \label{eq3}
\end{equation*}
where $\sigma=\pm 1$ for spin up/down, $G_\sigma=G_0(1+\sigma P)$ is the spin-dependent tunneling conductance, $N_\sigma=(N_0+\sigma N_z)/2$ is the spin-dependent density of states, $f_0(\epsilon)=[\exp(\epsilon/T)+1]^{-1}$ is the Fermi function, $G_0$ is the normal-state tunneling conductance, $N_{0/z}$ is the spin average/difference density of states, and $P\in [-1,1]$ is the spin polarization. Carrying out the sum over the spin yields
\begin{equation*}
    \label{eq:diode-eq-I}
    I(V) = G_0 \int d\epsilon [N_0+P N_z][f_0(\epsilon-V)-f_0(\epsilon)].
\end{equation*}
The distribution function factor can be simplified as
\begin{equation*}
\begin{split}
f(\epsilon-V)-f(\epsilon) = \frac{1}{e^{(\epsilon-V)/T}+1} - \frac{1}{e^{(\epsilon)/T}+1} \\
=\frac{1-e^{-V/T}}{1+e^{-V/T}+e^{-\epsilon/T}+e^{(\epsilon-V)/T}}.
\end{split}
\end{equation*}
Because of the gap in the $N_0$ and $N_z$ functions, this needs to be evaluated only for $\epsilon > \Delta-h$ and for $\epsilon < -\Delta+h$. If $V \ll \Delta-h$, for the positive energies the last term in the denominator is larger than the others so we may approximate
$$
f(\epsilon-V)-f(\epsilon) \approx (1-e^{-V/T})e^{-(\epsilon-V)/T}.
$$
On the other hand, for negative energies the third term in the denominator is larger than the others and we may approximate
$$
f(\epsilon-V)-f(\epsilon) \approx (1-e^{-V/T})e^{-\epsilon/T}.
$$
In the absence of spin relaxation or orbital depairing, the spin-dependent DoS is
\begin{align*}
N_0 + P N_z = & {\rm Re}\Bigg[\frac{|\epsilon+h]}{\sqrt{(\epsilon+h)^2-\Delta^2}} \frac{1+P}{2} \\ 
 &+ \frac{|\epsilon-h|}{\sqrt{(\epsilon-h)^2-\Delta^2}}\frac{1-P}{2}\Bigg],
\end{align*}
we get the current to the form
\begin{align*}
I=&\frac{G_0}{2}(1-e^{-V/T})\Bigg[\int_{\Delta-h}^\infty \frac{(\epsilon+h)(1+P)}{\sqrt{(\epsilon+h)^2-\Delta^2}} e^{-(\epsilon-V)/T} d\epsilon \\
& + \int_{\Delta+h}^\infty \frac{(\epsilon-h)(1-P)}{\sqrt{(\epsilon-h)^2-\Delta^2}}e^{-(\epsilon-V)/T} \\
& - \int_{-\infty}^{-\Delta-h}\frac{(\epsilon+h)(1+P)}{\sqrt{(\epsilon+h)^2-\Delta^2}} e^{\epsilon/T} d\epsilon \\ 
& - \int_{-\infty}^{-\Delta+h} \frac{(\epsilon-h)(1-P)}{\sqrt{(\epsilon-h)^2-\Delta^2}}e^{\epsilon/T}d\epsilon\Bigg]
\end{align*}
Shifting the energies by the spin-splitting field up and down, and reverting the sign of the energy in the latter two integrals yields
\begin{align*}
I=& \frac{G_0 (1-e^{-V/T})}{2}[(1+P)e^{(h+V)/T}+(1-P)e^{(V-h)/T} \\
& +(1+P)e^{-h/T} + (1-P)e^{h/T}] \times\underbrace{\int_{\Delta}^\infty \frac{\epsilon e^{-\epsilon/T}}{\sqrt{\epsilon^2-\Delta^2}}d\epsilon}_{=\Delta K_1(\Delta/T)},
\end{align*}
where $K_1(\Delta/T) \approx \sqrt{\pi/2} e^{-\Delta/T}\sqrt{\frac{T}{\Delta}}$ is the Bessel $K$ function and the latter approximation is valid for $\Delta \gg T$. Let us define
\begin{equation}
    I_S \equiv G_0 \Delta K_1\left(\frac{\Delta}{T}\right)e^{h/T}.
    \label{eq:Is}
\end{equation}
Now rearranging terms in the expression for the current allows us to write it as
\begin{equation}
    \begin{split}
    I(V) = & I_S\left(e^{V/T}-1\right)+ I_S e^{-2h/T}  (1-e^{-V/T}) \\
    & + I_S(1-e^{-2h/T}) \left[\cosh\left(\frac{V}{T}\right)-1\right](P-1).
    \label{eq:IVI}
\end{split}
\end{equation}
This also yields the rectification
\begin{equation}
    R = P \tanh\left(\frac{h}{T}\right) \tanh\left(\frac{V}{2T}\right).
\end{equation}
For $h\gg T$ we get Eq.~\eqref{eq:IV} and the corresponding simplified expression for $R$ quoted in the main text.
\newline
\newline
\textbf{Corrections to the current due to subgap states.}
Inelastic scattering introduces subgap states, which can be well described within the Dynes model \cite{dynes_tunneling_1984}. At low energies ($\epsilon<\Delta-h$), a weak Dynes parameter $\Gamma\ll \Delta-h$ introduces a correction to the superconducting density of states given as $\delta N_\sigma(\epsilon)=\frac{\Gamma}{\rho_\sigma}(1+\frac{\epsilon_\sigma^2}{\rho_\sigma^2})$. Here $\epsilon_\sigma=\epsilon+\sigma h$ and $\rho_\sigma=\sqrt{\Delta^2-\epsilon_\sigma^2}$. Combining this with Eq.~\eqref{eq3}, we find the following correction to the current, valid at low temperatures and for voltages $V<\Delta-h$:
\begin{equation}
\delta I=\Gamma G_0 [F_{asym}(eV,h)+P F_{sym}(eV,h)].
\end{equation}
Here we introduced the functions $F_{asym}(eV,h)=\frac{1}{2}[F(eV+h)+F(eV-h)]$, $F_{sym}(eV,h)=\frac{1}{2}[F(eV+h)-F(eV-h)-2F(h)]$, with $F(x)=x/\sqrt{\Delta^2-x^2}$. For small voltages and weak exchange field, $h,eV\ll \Delta$, we may approximate $\delta I \approx \frac{\Gamma eV}{\Delta}[1+\frac{3}{2}\frac{eVh}{\Delta^2}]$.

Taking into account the correction $\delta I$ together with Eq.~\eqref{eq:IV}, the expression for the rectification coefficient $R$ becomes ($h\gg k_B T$)
\begin{equation}
R=P\frac{2 \sinh^2\frac{eV}{2k_BT}+ \xi F_{sym}(eV,h)}{ \sinh \frac{eV}{k_BT}+\xi F_{asym}(eV,h)},
\end{equation}
where $\xi=\frac{G_0\Gamma}{I_S}\sim \frac{\Gamma}{\Delta}e^{\Delta/(k_BT)}$. If the temperature is high-enough,  $k_BT\gg \Delta/\log(\frac{\Delta}{\Gamma})$, we have $\xi\ll 1$, and inelastic scattering can be neglected. In this case we obtain the expression shown in the main text: $R=P \tanh[eV/(2k_BT)]$. However, in the opposite  regime of very low temperatures, $k_B T\ll \Delta/\log(\frac{\Delta}{\Gamma})$, we find that $\delta I$ actually provides the dominant contribution to the current. In that case, $R=P F_{sym}/F_{asym}.$ Note that in both regimes the maximal rectification coefficient is given by $R_{max}=P$. In the first regime, the maximum is reached at $eV \sim k_B T$, whereas in the second it is at $eV\sim \Delta-h$.    
\newline
\newline
\textbf{Model for the density of states (DoS).} 
In the calculation of the $I(V)$ characteristics the density of states of the superconductor, $\mathcal{N}_\sigma(\epsilon)$, plays a central role. 
We obtain it from the quasiclassical Green's functions (GFs), $\check g$, defined in the Nambu$\otimes$spin space. 
These are $4\times 4$ matrices that satisfy the normalization condition, $\check g^2 = 1$.
Here the "check" symbol, $\check \cdot$, indicates $4\times 4$ matrices.

In the bulk of a dirty superconductor with a constant exchange field aligned along a given axis, the \textit{retarded} quasiclassical GFs fulfill the following Usadel equation\cite{usadel_generalized_1970,rouco2019charge,heikkila2019thermal}:
\begin{equation}
\label{eq:usadel}
    \big[i(\epsilon + i\Gamma) \hat\tau_3 + ih \hat\tau_3\hat\sigma_z - \Delta\hat\tau_1 - \check\Sigma,
    \check g\big] = 0.
\end{equation}
Here, $\epsilon$ is the energy, $\Gamma$ is a small energy term known as the Dynes parameter\cite{dynes_tunneling_1984}, $h$ stands for the strength of the exchange field, $\Delta$ is the self-consistent superconducting order parameter and $\hat\tau_i$ and $\hat\sigma_a$ label the Pauli matrices spanning Nambu and spin space, respectively. 
Direct product between Pauli matrices spanning different spaces is implied, and identity matrices, $\hat\tau_0$ and $\hat\sigma_0$, are obviated.
The square brackets, $[\cdot,\cdot]$, stand for commutation operation and $2\times2$ matrices are indicated with a $\hat \cdot$ symbol. 
A typical value of the Dynes parameter is $\Gamma \sim 10^{-3}\Delta$ and its importance is twofold: first it avoids analytical problems in the computation of the GFs and second it models the effect of non-elastic processes in the superconductor.
The $\check \Sigma$ matrix is the self-energy that consists of three contributions:
\begin{equation}
    \check \Sigma = \check \Sigma_{so} + \check \Sigma_{sf} + \check \Sigma_{orb}.
\end{equation}
From left to right, these are the spin relaxation due to spin-orbit coupling, the spin relaxation due to spin-flip events and the orbital depairing due to external magnetic fields, respectively. Explicitly, each contribution within the relaxation time approximation, reads:
\begin{equation}
    \check\Sigma_{so} = \frac{\hat\sigma_a \check g \hat\sigma_a}{8\tau_{so}},
    \qquad
    \check\Sigma_{sf} = \frac{\hat\sigma_a \hat\tau_3 \check g \hat\tau_3 \hat\sigma_a}{8\tau_{sf}}, 
    \qquad
    \check\Sigma_{orb} = \frac{\hat\tau_3 \check g \hat\tau_3}{\tau_{orb}}.
\end{equation}
Here $\tau_{so}$, $\tau_{sf}$ and $\tau_{orb}$ stand for spin-orbit, spin-flip and orbital depairing relaxation times, respectively, and we sum over repeated indices.
We estimate the orbital depairing in the superconducting layer due to an applied magnetic field as\cite{anthore-2003-density,de_gennes_superconductivity_1966}:
\begin{equation}
    \tau_{orb}^{-1} = \left(\frac{\pi d \xi_0 B}{\sqrt{6} \Phi_0}\right)^2 \Delta_0,
\end{equation}
where $\Phi_0$ is the quantum of magnetic flux, $d$ stands for the width of the superconducting layer, $B$ is the applied magnetic field, $\Delta_0$ is the superconducting gap at zero field ($T=0$ and $h=0$) and $\xi_0$ is the superconducting coherence length. 

In addition to Eq.~\eqref{eq:usadel}, the value of the superconducting gap is related to the quasiclassical GFs via the self-consistent equation,
\begin{equation}
    \Delta = \frac{\lambda}{8i} \int_{-\Omega_D}^{\Omega_D} d\epsilon \text{Tr}\big[\hat\tau_1 \check g\big],
    \label{eq:self-consistent-delta}
\end{equation}
where the trace runs over the Nambu$\otimes$spin space, $\lambda$ is the coupling constant and $\Omega_D$ is the Debye cutoff energy. 

From Eqs.~\eqref{eq:usadel}, \eqref{eq:self-consistent-delta} and the normalization condition we compute the value of $\check g$, from which the the spin average/difference density of states, $\mathcal{N}_{0/z}$, can be directly calculated:
\begin{equation}
    \mathcal{N}_{0/z} (\epsilon) = \frac{1}{2} \text{Re}\Big[\text{Tr}(\hat\tau_3 \hat\sigma_{0/z} \; \check g )\Big].
    \label{eq:DOS-0z}
\end{equation}
By fitting the experimental $I(V)$ curves with Eqs.~\eqref{eq:diode-eq-I} and \eqref{eq:DOS-0z} we are able to obtain the different parameters used in the model.  
\newline
\newline
\textbf{Model for transverse rectification.}
In Fig.~\ref{fig3}(c), we calculate the rectification voltage from the experimentally measured $I(V)$ curves at different heating currents $I_H$ using the following theoretical model. The open circuit voltage $V_s$ for the transverse rectifier configuration shown in Fig.~\ref{fig3}(a) can be determined by solving the equation
\begin{equation}
\int_0^1 I(s I_H R_x+V_s+V_{inst})ds=0.
\end{equation}
Here, $R_x$ is the lateral resistance of the junction, and $V_{inst}$ is the instrumental offset, which is obtained from $I(V_{inst})=0$ at $I_H=0$. The voltage $V_s$ contains two contributions: a larger trivial Ohmic contribution due to the heating current, and a smaller contribution due to the rectification effect. The former is antisymmetric in $I_H$, whereas the latter is symmetric. Therefore, the symmetrized voltage
\begin{equation}
V_{\rm sym}=\frac{1}{2}[V_s(I_H)+V_s(-I_H)]
\end{equation}
comes from the rectification effect only. 

%\clearpage
\bibliography{refs.bib}

\section*{Acknowledgement}
This work was mainly supported by the EU’s Horizon 2020 research and innovation program under Grant Agreement No. 800923 (SUPERTED).
E.S. and F.G. acknowledge the European Research Council under Grant Agreement No. 899315 (TERASEC), and  the  EU’s  Horizon 2020 research and innovation program under Grant Agreement No. 964398 (SUPERGATE) for partial financial support.
 M.S. and E.S. acknowledge partial funding from the European Union’s Horizon 2020 research and innovation programme under the Marie Sk\l{}odowska Curie Action IF Grant No. 101022473 (SuperCONtacts).
J.M.  acknowledges financial support in the USA by the Army Research Office (grant ARO W911NF-20-2-0061), ONR (grant N00014-20-1-2306), NSF (grant DMR 1700137) and NSF C-Accel Track C Grant No. 2040620.

\section*{Data availability} The data that support the findings of this study are available within the paper and its Supplementary Information. Source data are provided with this paper.

\section*{Author contribution}
E.S., N.L. and M.S. performed the experiments and analyzed the data.
S. I., M.R, P.V., T.H., and F.S.B. provided theoretical support.
J.M. fabricated the Cu/EuS/Al devices and C.G.O, M.I and C.R. the EuS/Al/AlO$_{x}$/Co ones.
E.S. conceived the experiment together with F.G. who supervised the project.
E.S. and M.S. wrote the manuscript with feedback from all authors.

\section*{Competing interests}

The authors declare the following competing interests: with the Instituto Nanoscienze-CNR, the following authors: E.S., M.S., F.G., P.V., T.T.H., S.I and F.S.B., have filed a patent (International Application N. PCT/IT2021/000038 "APPARATUS AND METHOD FOR SUPERCONDUCTING DIODE", status: pending, aspect of manuscript covered in patent application: rectification and diode-behavior of the material combinations, architecture and measurement configurations presented).

\section*{Correspondence}
Correspondence should be addressed to E.S. and M.S.

%----------------Supplementary Information-----------------
\clearpage
\onecolumngrid
\begin{center}
        \vspace*{0.5cm}
        \Large
        \textbf{Supplementary Information}
        \vspace{0.5cm}
\end{center}

\renewcommand{\thefigure}{S\arabic{figure}}
\renewcommand{\theequation}{S\arabic{equation}}
\setcounter{figure}{0}

%---------------------- Current mode---------------------
\section{Analysis of the diode in current bias.} 
\begin{figure*}[ht]
\centering
\includegraphics[scale=0.75]{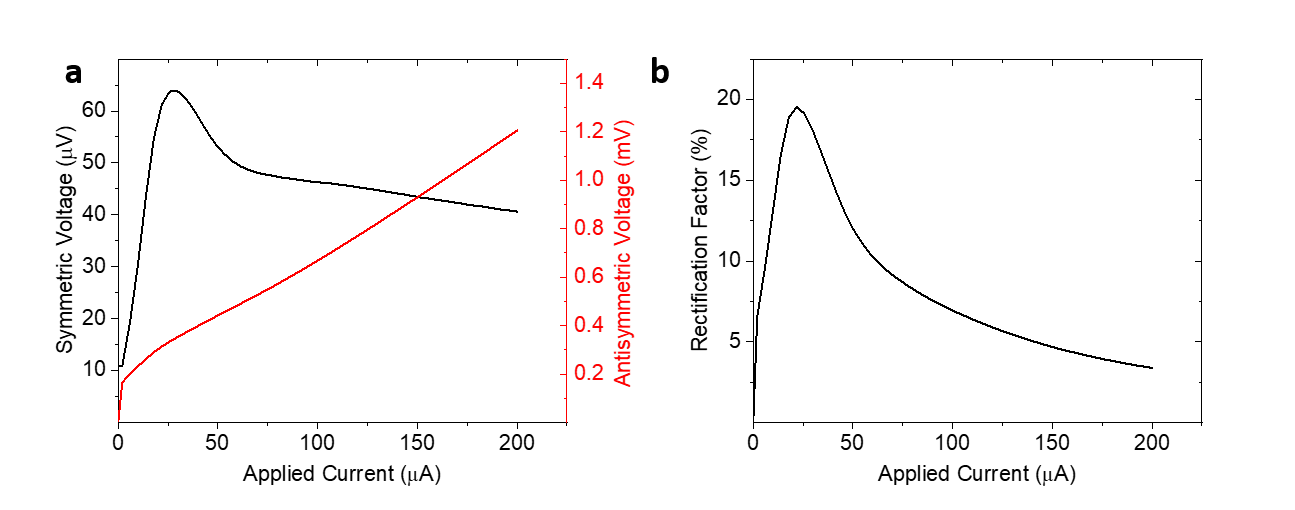}
\caption{
\textbf{a}, Symmetric and anti-symmetric parts ($V_{Sym}=(V(I)+V(-I))/2$, $V_{Antisym}=(V(I)-V(-I))/2$) of the $V(I)$ characteristic equivalent to the one shown in Fig.~2b of the main text. \textbf{b}, Rectification factor ($R=V_{Sym}/V_{Antisym}$) evaluated for this configuration. The maximum rectification occurs at low currents and reaches around $\sim 20 \%$ there.
 }
\label{figS1}
\end{figure*}

Starting from the $I(V)$ characteristics shown in Fig.~2b of the main text, it is possible to quantify the rectification efficiency when operating the diode in current bias. Notably, due to the anti-symmetric, non-linear response of the diode, this is not simply equivalent to the voltage bias analysis shown in the main text as one can see comparing it with Fig.~\ref{figS1}. In fact, the high resistance of the tunnel junction at low voltages promotes a sudden response of the anti-symmetric voltage (red curve in Fig.~\ref{figS1}a) while the symmetric component grows at larger current bias (black curve in Fig.~\ref{figS1}a). As a result, the rectification in current bias is smaller, with a maximum rectification of $\sim 20 \%$ as shown in Fig.~\ref{figS1}b.

%---------------------- Max's sample -----------------------
\section{Additional sample structures.}

Besides the superconductor/ferromagnetic insulator/normal metal (S/FI/N) structure shown in the main text, different material combinations with equivalent spin-filtering and spin-splitting have been investigated. Most notably, a FI/S/I/F structure (where I is an insulator and F is a metallic ferromagnet) have been investigated. Differing from S/FI/N junctions, here, the spin-filtering and spin-splitting are decoupled. The former is still provided by the FI/S interface, while the latter is due to the I/F tunnel barrier. 

Samples are cross-bars made by electron-beam evaporation employing an in-situ shadow mask on a substrate of fused silica and consist of layers of EuS(14nm)/ Al(9nm)/ AlO$_{x}$(4-5nm)/ Co(10nm)/ CaF(7nm). The overlap between the Al and the Co strip has an area of $~300\times300$ $\mu$m$^2$. The tunneling spectroscopy is carried out at cryogenic temperatures down to 50 mK in a filtered cryogen-free dilution refrigerator. The $I(V)$ characteristics are obtained from DC four-wire measurements as described in the main text.

\begin{figure*}[ht]
\centering
\includegraphics[scale=0.65]{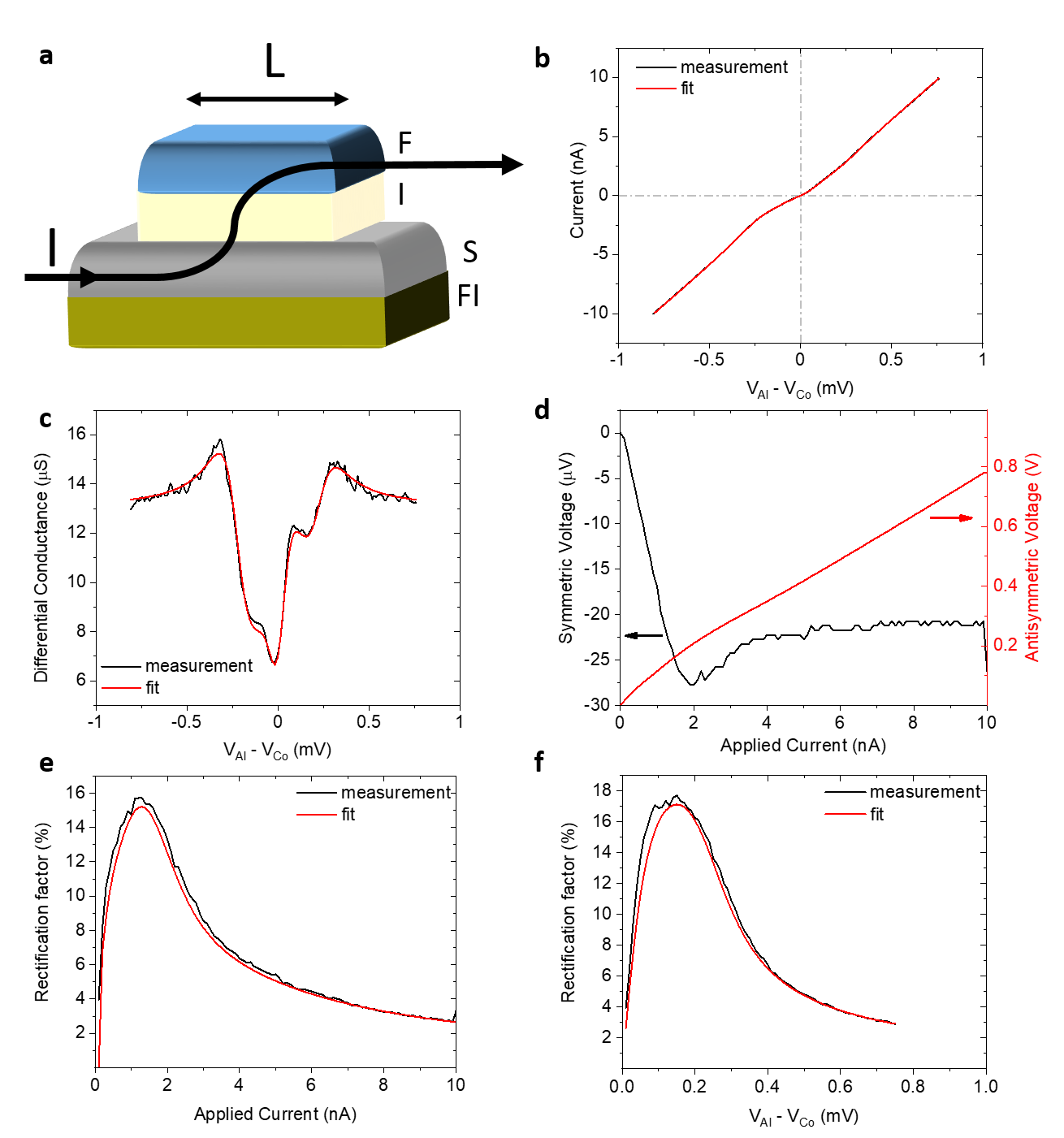}
\caption{\textbf{Rectification of a superconducting diode made with alternative materials: a FI/S/I/F junction.}
\textbf{a}, Schematic of the tunnel junction. The path of the tunneling current is indicated by the black line and its arrows. 
\textbf{b}, Current-to-Voltage ($I(V)$) characteristic of the junction measured at $T\simeq \SI{50}{\milli\kelvin}$, $B= \SI{0}{\tesla}$.
\textbf{c}, Differential conductance obtained from the numerical derivative of \textbf{b}. 
The fits in \textbf{b} and \textbf{c} have been obtained form Eq.~(3) and (14) of the main text with the following  parameters: $\Delta_0 = 0.228$ meV, $h=0.097$ meV, $P=0.3$, $\Gamma=0.01$ meV, $T=250$ mK, $\tau_{sf}=15$ meV$^{-1}$, $\tau_{so}=600$ meV$^{-1}$.
\textbf{d}, Symmetric and anti-symmetric parts of the $I(V)$ characteristic in \textbf{b} showing a sizable symmetric component of the voltage.
\textbf{e} and \textbf{f}, Rectification coefficients evaluated from the $I(V)$ characteristics in current and voltage bias, respectively (black line). The comparison of the rectification extracted from the full numerical model is also shown (red line). 
} 
\label{figS2}
\end{figure*}

The data analysis on the $I(V)$ characteristic at $B=0$ is shown in Fig.~\ref{figS2}. Notably, as shown in Fig.~\ref{figS2}a and b, in this device the zero bias conductance is more pronounced with  respect to the the S/FI/N sample shown in the  main text. On the other hand, large spin-splitting and spin-filtering is visible even at zero magnetic field thanks to the stronger ferromagnetism of the EuS layer. Therefore, even if the rectification is smaller with respect to the S/FI/N devices (here the maximum rectification is $\sim 18 \%$ as estimate in Fig.~\ref{figS2}d and f) the presence of a sizable rectification, even in the absence of an external magnetic field, makes it appealing for applications. Moreover, differing from S/FI/N junctions where the direction of the diode is fixed by the sign of the exchange interactions at the EuS/Al interface, in this typology of device the direction of the diode can be inverted by changing the relative magnetization of the FI and F layers (parallel or anti-parallel), introducing additional functionalities.

%--------------FIGS2--------------

%---------------------- NEP discussion ----------------------
%\newpage
\section{Applications for detection.}
The sizable rectification of the superconducting tunnel diode observed both in the direct (i) and in the transverse (ii) configuration can find an immediate application in the detection of electromagnetic radiations. Starting from the characterizations presented in the main text, it is possible to estimate the maximum resolution and the noise equivalent power (NEP) of a detector based on this technology. For configuration (i) the DC response to a sinusoidal AC signal ($V_{AC}=V_0 \sin{(\omega t)}$) can be estimated by averaging the current response over the signal time period T:

\begin{equation}
\label{Idc}
I_{DC}=\int_{-T/2}^{T/2} I(V_{AC}) \, \frac{dt}{T} = \int_{-T/2}^{T/2} I_{Sym}(V_{AC}) \, \frac{dt}{T},
\end{equation}
and the resulting power dissipated by the signal reads: 
\begin{equation}
\label{PIdc}
P=\int_{-T/2}^{T/2} V_{AC}  \, I(V_{AC}) \, \frac{dt}{T} = \int_{-T/2}^{T/2} V_{AC} \, I_{Antisym}(V_{AC}) \, \frac{dt}{T}.
\end{equation}

\begin{figure*}[ht]
\centering
\includegraphics[scale=0.8]{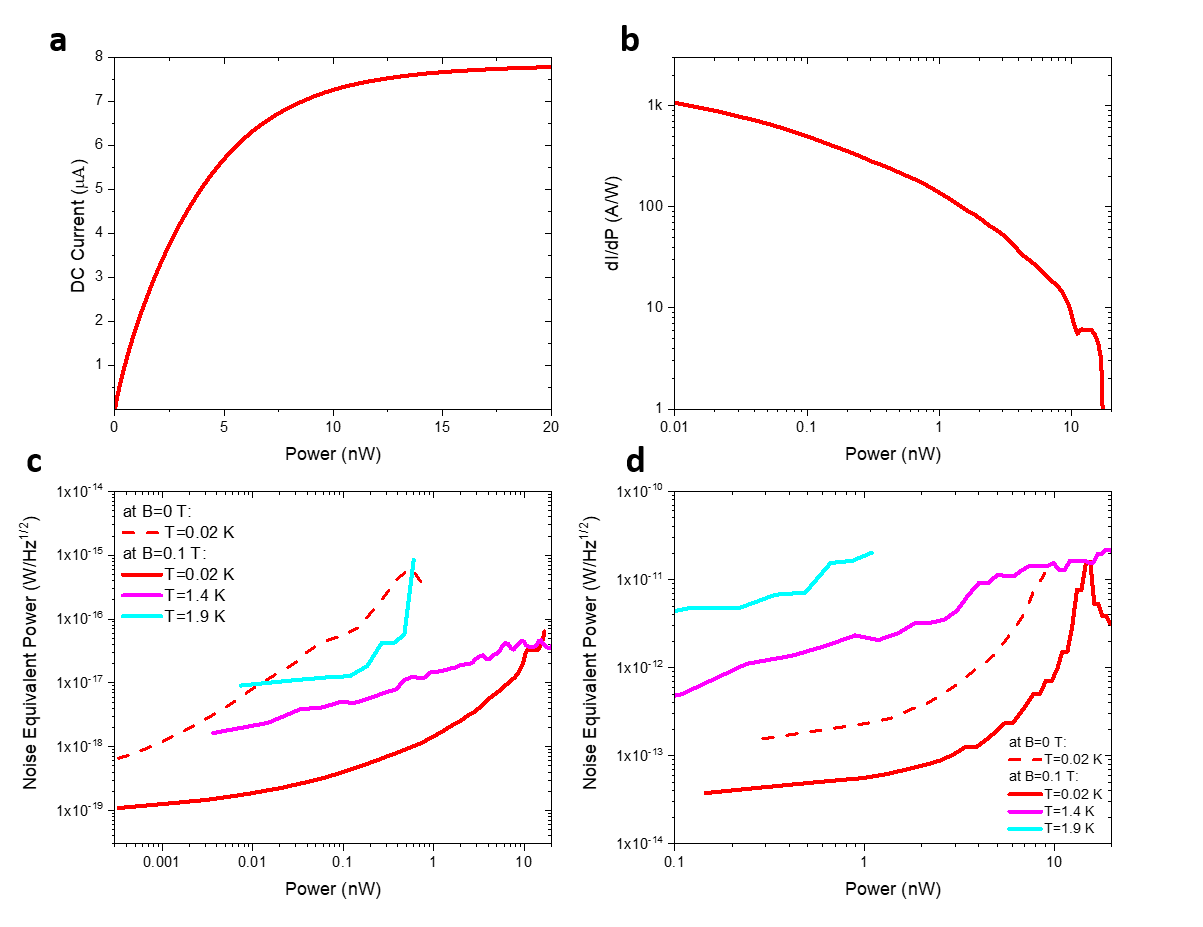}
\caption{\textbf{Resolution and NEP of the N/FI/S junction in the direct configuration.}
\newline
\textbf{a}, DC rectified current ($I_{DC}$) vs. the power of the input AC signal evaluated from the $I(V)$ characteristic in Fig.~2b and Eq.~\ref{Idc}, \ref{PIdc} in the closed circuit configuration at 20~mK with an external magnetic field of 0.1~T.
\textbf{b}, Power spectrum of the transfer function ($dI/dP$) resulting from \textbf{a}.
\textbf{c}, Power spectrum of the NEP evaluated by the ratio between the input noise of a room temperature current amplifier (FEMTO LCA-2-10T) $\sim \SI[per-mode=symbol]{0.2}{\femto\ampere\per\sqrt{\hertz}}$ and the transfer function in \textbf{b}. The estimations have been done for different magnetic fields and temperatures as indicated in the legend.
\textbf{d}, NEP evaluated for the open circuit configuration, starting from Eq.~\ref{Vdc} and ~\ref{PVdc}, with an input noise voltage of $\sim \SI[per-mode=symbol]{0.4}{\nano\volt\per\sqrt{\hertz}}$  (amplifier: DLPVA-100-BUN-S). The estimations have been done for different magnetic fields and temperatures as indicated in the legend.}
\label{figS3}
\end{figure*}

In Fig.~\ref{figS3}a we show the $I_{DC}(P)$ estimated from the latter equations applied to the  $I(V)$ characteristic of the superconducting diode in the direct configuration presented in Fig.~2b of the main text. The resulting transfer function  ($dI_{DC}/dP$) is shown in Fig.~\ref{figS3}b and determines the resolution of the detector. If terminated with a low-noise current amplifier with an input noise of $\sim \SI[per-mode=symbol]{0.2}{\femto\ampere\per\sqrt{\hertz}}$ 
(e.g. FEMTO LCA-2-10T as a conventional room-temperature amplifier) the detector can provide a very low NEP down to 
$\sim \SI[per-mode=symbol]{1e-19}{\watt\per\sqrt{\hertz}}$ at low powers (see red solid line in Fig.\ref{figS3}c for the full power spectrum at 0.1 T and 0.02 K), already competing with state-of-the-art detectors. A high sensitivity is, however, preserved even at zero magnetic fields, with a NEP of $\sim \SI[per-mode=symbol]{1e-18}{\watt\per\sqrt{\hertz}}$ at low power (<pW) increasing up to $\sim 10^{-16}$ at few nW (see red dashed line in Fig.\ref{figS3}c). A similar monotonic degradation applies to higher temperatures (up to 1.9 K) at 0.1 T with a NEP of $10^{-17}$ $\sim \SI[per-mode=symbol]{1}{\watt\per\sqrt{\hertz}}$ (see cyan and magenta lines in Fig.\ref{figS3}c).

Fig.\ref{figS3}d shows the same evaluations in an open circuit configuration where the rectified DC voltage and power are quantified in a similar mode:

\begin{equation}
\label{Vdc}
V_{DC}=\int_{-T/2}^{T/2} V(I_{AC}) \, \frac{dt}{T} = \int_{-T/2}^{T/2} V_{Sym}(I_{AC}) \, \frac{dt}{T},
\end{equation}

\begin{equation}
\label{PVdc}
P=\int_{-T/2}^{T/2} I_{AC}  \, V(I_{AC}) \, \frac{dt}{T} = \int_{-T/2}^{T/2} I_{AC} \, V_{Antisym}(I_{AC}) \, \frac{dt}{T}.
\end{equation}

where $I_{AC}=I_0 \sin{(\omega t)}$ is the AC signal to probe.
Here, the resulting NEP evaluated with an input noise voltage of $\sim \SI[per-mode=symbol]{0.4}{\nano\volt\per\sqrt{\hertz}}$ is significantly smaller ($\sim \SI[per-mode=symbol]{1e-14}{\watt\per\sqrt{\hertz}}$), probably due to the small impedance of the tunnel junction favoring the detection in the closed circuit configuration. 
 
In the transverse configuration (ii), the rectification response can be estimated from Eq.~\ref{Vdc} in a similar way, starting from the $V_{sym}(I_{Cu})$ characteristics shown in Fig.~3 and Fig.~4 of the main text. Differing from configuration (i), the power will be mainly dissipated in the Cu strip and can be estimated from its simple Ohmic response:

\begin{equation}
P=\int_{-T/2}^{T/2} I_{AC}^2 \, R \, \frac{dt}{T} = \frac{I_0^2 \,R}{2},
\end{equation}
where $R\simeq 2  \Omega$ is the lateral resistance of the Cu lead at the interface with the EuS.

\begin{figure*}[ht]
\centering
\includegraphics[scale=0.75]{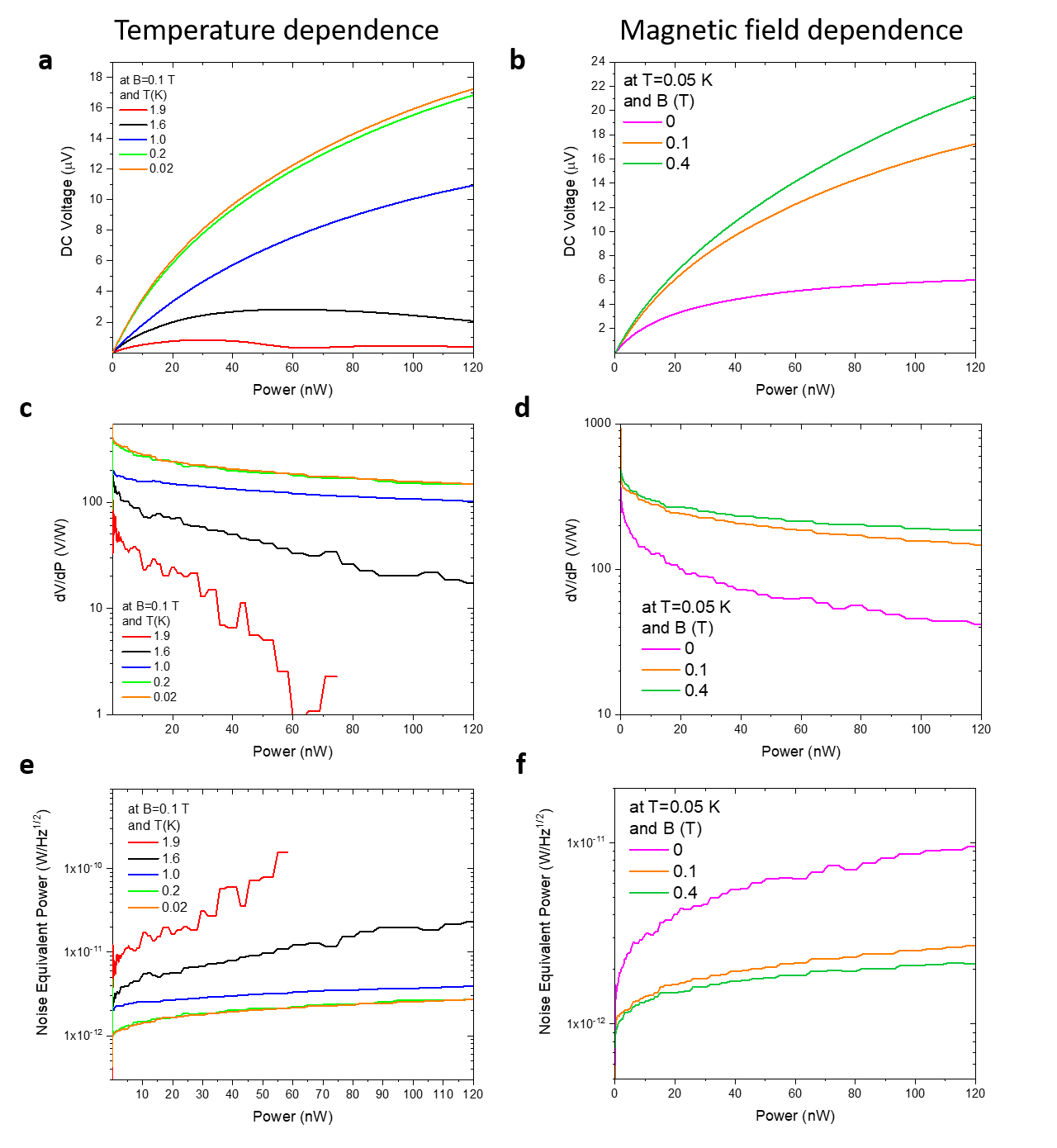}
\caption{\textbf{Resolution and NEP of the N/FI/S junction in the transverse configuration.}
\textbf{a}, DC rectified voltage vs. the power of the input AC signal estimated in the transverse open-circuit configuration at different temperatures and \textbf{b}, magnetic fields.
\textbf{c}, Power spectrum of the transfer function ($dV/dP$) resulting from \textbf{a}. The equivalent power spectrum resulting from \textbf{b} is shown in \textbf{d}. 
In panel \textbf{e}, the power spectrum of the NEP evaluated from \textbf{c} is shown for an input noise voltage of $\sim \SI[per-mode=symbol]{0.4}{\nano\volt\per\sqrt{\hertz}}$  (amplifier: DLPVA-100-BUN-S). The equivalent power spectrum resulting from \textbf{d} is shown in \textbf{f}.
}
\label{figS4}
\end{figure*}

In Fig.~\ref{figS4}a and d, the  $V_{DC}(P)$ is shown for different temperatures and magnetic fields, respectively. The resulting resolutions $dV_{DC}/dP$ are shown in  Fig.~\ref{figS4}b and e. The NEP estimated with an input noise of 
$\sim \SI[per-mode=symbol]{0.4}{\nano\volt\per\sqrt{\hertz}}$ 
 (e.g. DLPVA-
100-BUN-S as a room-temperature voltage amplifier) is shown in Fig.~\ref{figS4}c and f.
Notably, in this configuration the NEP of $\sim$ $10^{-11}$ - $10^{-12}$ $\SI[per-mode=symbol]{}{\watt\per\sqrt{\hertz}}$ is one order of magnitude larger with respect to the NEP evaluated in configuration (i) for the same open circuit configuration (Fig.~\ref{figS3}d). On the other hand, in this configuration the detector is sensitive to a larger range of powers (up to 120~nW with no sign of saturation) and the DC signal originated across the junction is already decoupled from the AC component flowing in the Cu strip.

\begin{figure*}[ht]
\centering
\includegraphics[scale=0.75]{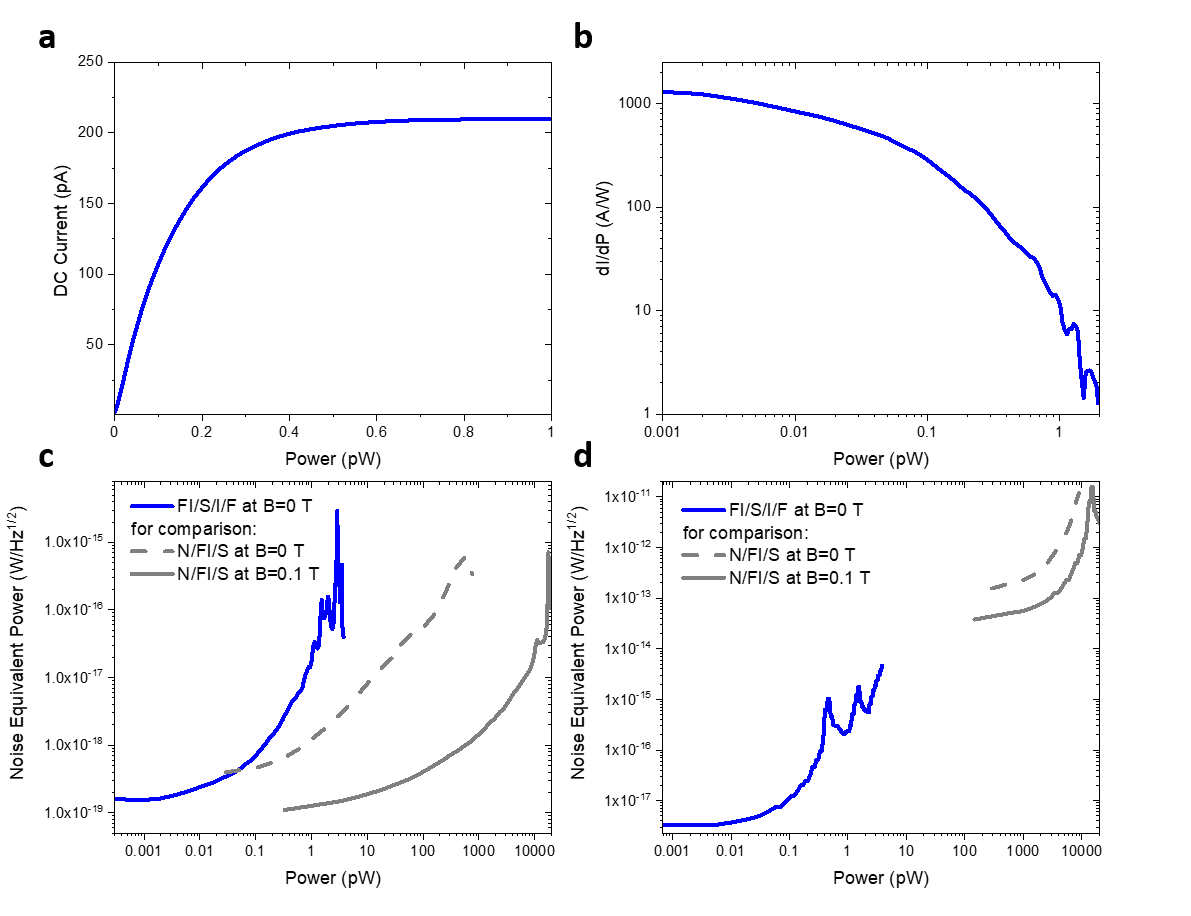}
\caption{\textbf{Resolution and NEP of the FI/S/I/F junction in the direct configuration.}
\textbf{a}, DC rectified current ($I_{DC}$) vs. the power of the input AC signal evaluated from the $I(V)$ characteristic in Fig.~\ref{figS2}b and Eq.~\ref{Idc}, \ref{PIdc} in the closed circuit configuration at 50~mK and zero magnetic field.
\textbf{b}, Power spectrum of the transfer function ($dI/dP$) resulting form \textbf{a}.
\textbf{c}, Power spectrum of the NEP evaluated by the ratio between the input noise of a room temperature current amplifier (FEMTO LCA-2-10T) $\sim \SI[per-mode=symbol]{0.2}{\femto\ampere\per\sqrt{\hertz}}$ and the transfer function in \textbf{b}. A comparison with the NEP evaluated for the N/FI/S junction is also shown (gray lines)
\textbf{d}, NEP evaluated for the open circuit configuration, starting from Eq.~\ref{Vdc} and ~\ref{PVdc}, with an input noise voltage of $\sim \SI[per-mode=symbol]{0.4}{\nano\volt\per\sqrt{\hertz}}$  (amplifier: DLPVA-100-BUN-S). A comparison with the NEP evaluated for the N/FI/S junction is also shown (gray lines).}
\label{figS5}
\end{figure*}

In Fig.~\ref{figS5} we estimate the resolution and NEP in direct configuration (i) for the additional sample structure FI/S/I/F introduced earlier ($I(V)$ shown in Fig.~\ref{figS2}). Thanks to the strong ferromagnetism of this device even at no applied external magnetic field the NEP reaches an impressive $\sim$ $10^{-18}$ - $10^{-19}$ $\SI[per-mode=symbol]{}{\watt\per\sqrt{\hertz}}$, but only for low powers due to the higher impedance (four orders of magnitude) of the tunnel junction. 
Such a high impedance improves the NEP in the closed circuit configuration reaching values of  $\sim$ $10^{-17}$ - $10^{-18}$ $\SI[per-mode=symbol]{}{\watt\per\sqrt{\hertz}}$, which is much smaller then the N/FI/S counterpart.
%--------------FIGS3----------------------

%----------FIGS4----------------

%-------------FIGS5--------------------

%\newpage
\section{Modeling contributions of rectification and thermovoltage.} 

To obtain input parameters for the modeling including the thermoelectric effects, we have fitted the IV data sets with the following model:
\begin{align}
    \frac{dI_{\rm{expt}}}{dV}(V_i,V_{H,j})
    \sim
    G_T
    \frac{d\tilde{I}_{\rm model}}{dV}(aV_{H,j}, V_i + V_{\mathrm{off},j}, T_{N,j})
    \,,
\end{align}
where $G_T$, $a$, $V_{\mathrm{off},j}$ and $T_{N,j}$ are the fit parameters, corresponding to a set of values $V_i$ and $V_{H,j}$ for the bias and heating voltages, and $dI_{\rm expt}/dV$ the observed differential conductance. The lateral resistance is $R_x=aR_H$, where $R_H\approx4.2\,\mathrm{k\Omega}$ is the resistance relating the heating voltage to the heating current, $I_H=V_H/R_H$. The theoretical current model is
\begin{align}
    \label{eq:Imodel}
    \frac{d\tilde{I}_{\rm model}}{dV}(V',V,T_N)
    =
    \int_{-1/2}^{1/2}ds\,
    G_{T,0}^{-1}\frac{dI_{NFIS}}{dV}(V + s V', T_N, T_S)
    \,,
\end{align}
where $I_{NFIS}(V,T_N,T_S)$ is the current-voltage relation discussed in Ref.~\cite{ozaeta_predicted_2014}. We include the effects of $\Gamma$ and other parameters affecting the density of states of the superconductor as in the main text, determined by separate fits done for $V_H=0$. We assume the order parameter $\Delta$ remains roughly constant in the parameter range considered, in which case the differential conductance is independent of the superconductor temperature $T_S$.

\begin{figure*}[ht]
\centering
\includegraphics[width=0.6\textwidth]{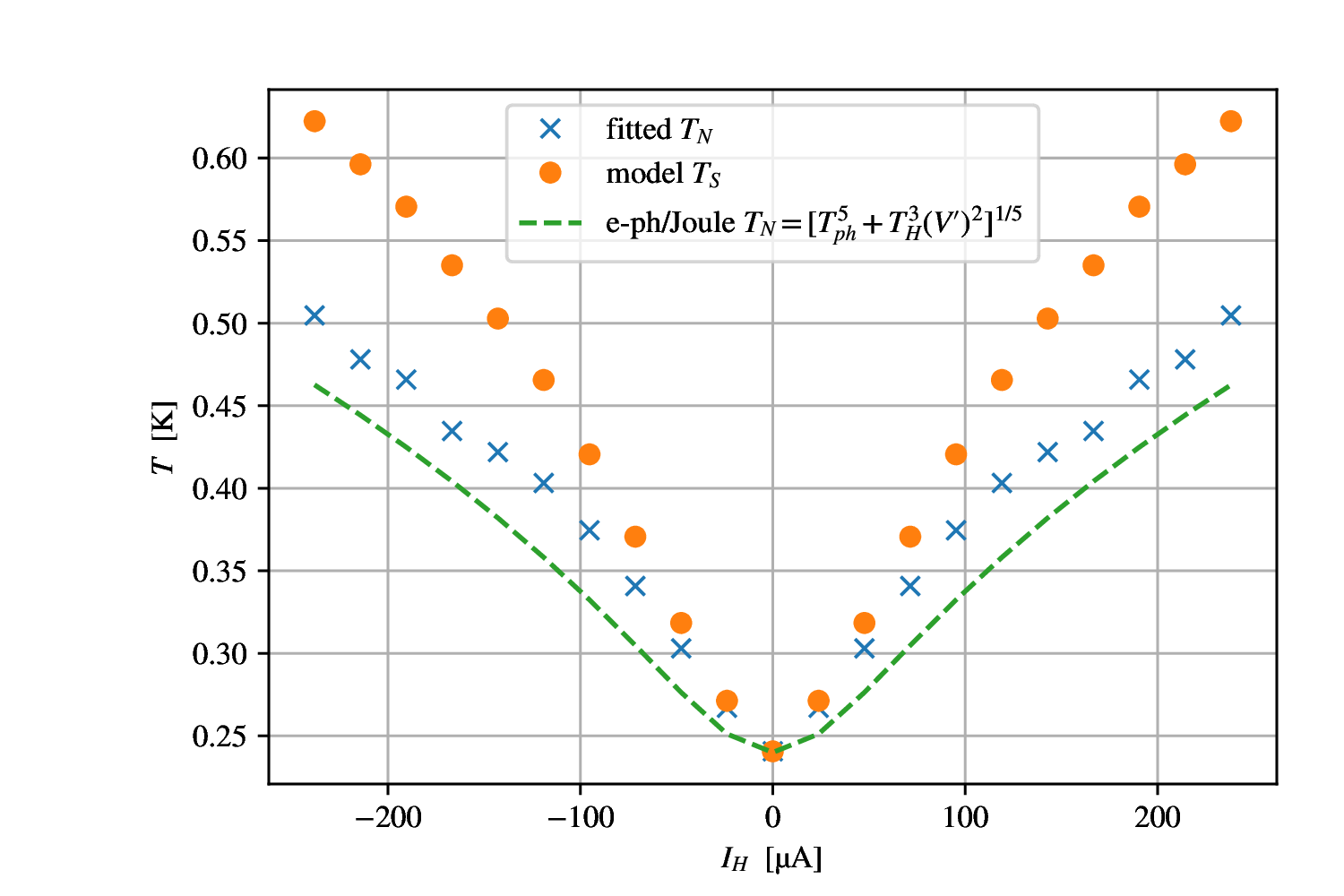}
\caption{
  \label{fig:TNvsIH}
  Temperature of the normal side $T_N$ obtained by fits to $dI/dV$
  measurements, and the temperature $T_S$ obtained from
  Eq.~\eqref{eq:Sheatbalance}.  Result from a local electron-phonon
  vs. Joule heating model for $T_N$, $\Sigma{}(T_N^5-T_{\rm ph}^5)=\rho
  j_H^2$, is also shown for comparison.
}
\end{figure*}

After obtaining the above parameters, we find the temperature $T_S$ of the superconducting side by solving the thermal balance model
\begin{align}
    \label{eq:Sheatbalance}
    \dot{Q}_{\rm tun}(V,T_N,T_S)
    =
    \dot{Q}_{\rm e-ph}(T_N,T_S)
    \,,
\end{align}
where $\dot{Q}_{\rm tun}$ is the tunneling heat current to $S$ obtained analogously as in Eq.~\eqref{eq:Imodel} (see Ref.~\cite{ozaeta_predicted_2014}). It is balanced by electron-phonon relaxation, with heat current $\dot{Q}_{\rm e-ph}$ as described in Ref.~\cite{heikkila2019thermal}, using literature parameters for Aluminum electron-phonon coupling \cite{giazotto_opportunities_2006}, and including the effects from spin splitting, $\Gamma$ and spin-flip scattering. The resulting $T_S$ is shown in Fig.~\ref{fig:TNvsIH}.

\begin{figure*}[ht]
\centering
\includegraphics[width=0.6\textwidth]{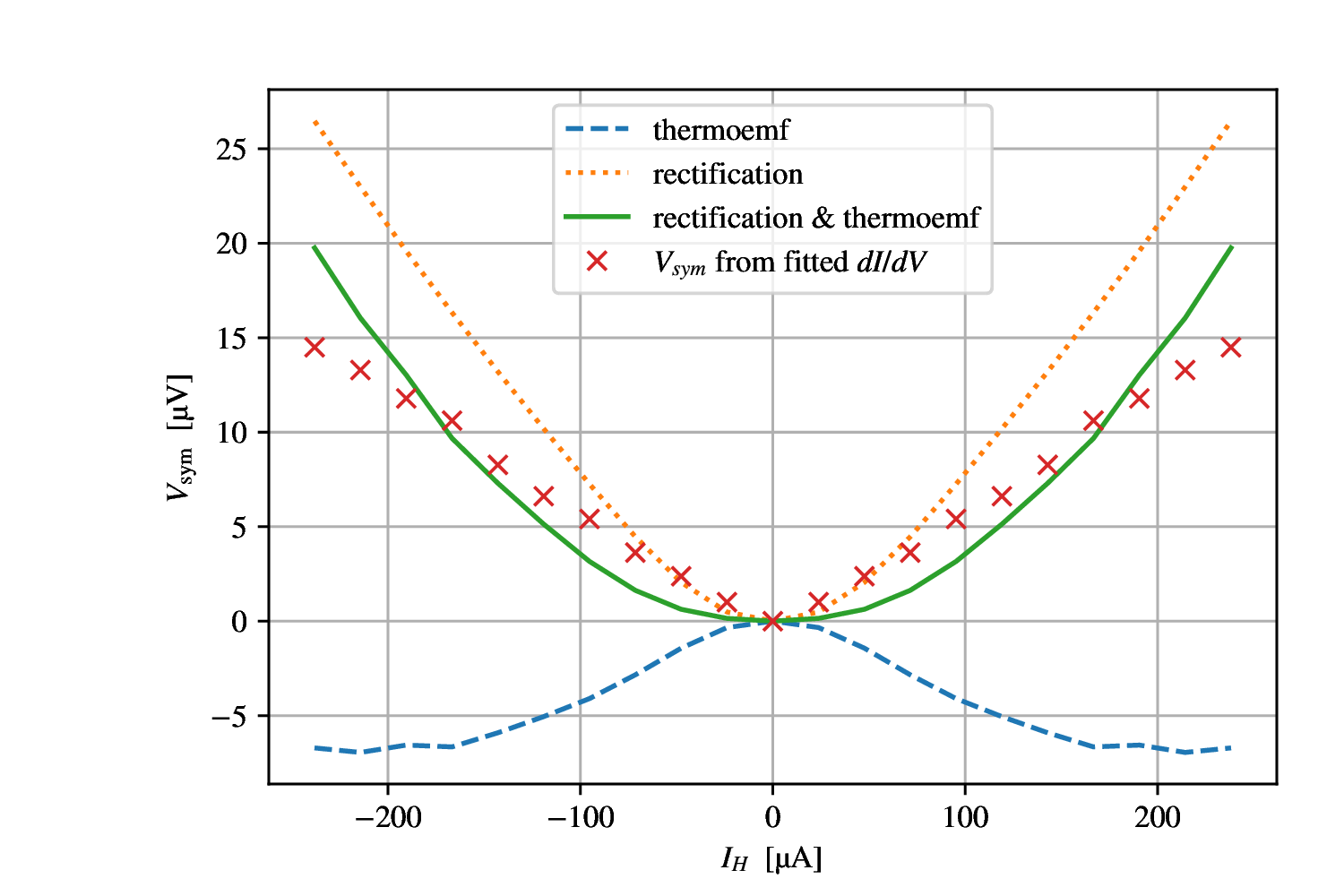}
\caption{
   \label{fig:Sthermoel}
   Symmetric part of the open-circuit voltage $V=V_N-V_S$ in the superconductor, modeled based on the $IV$ data at $B=200\,\mathrm{mT}$.
}
\end{figure*} 

\begin{figure*}[ht]
\centering
\includegraphics[width=0.6\textwidth]{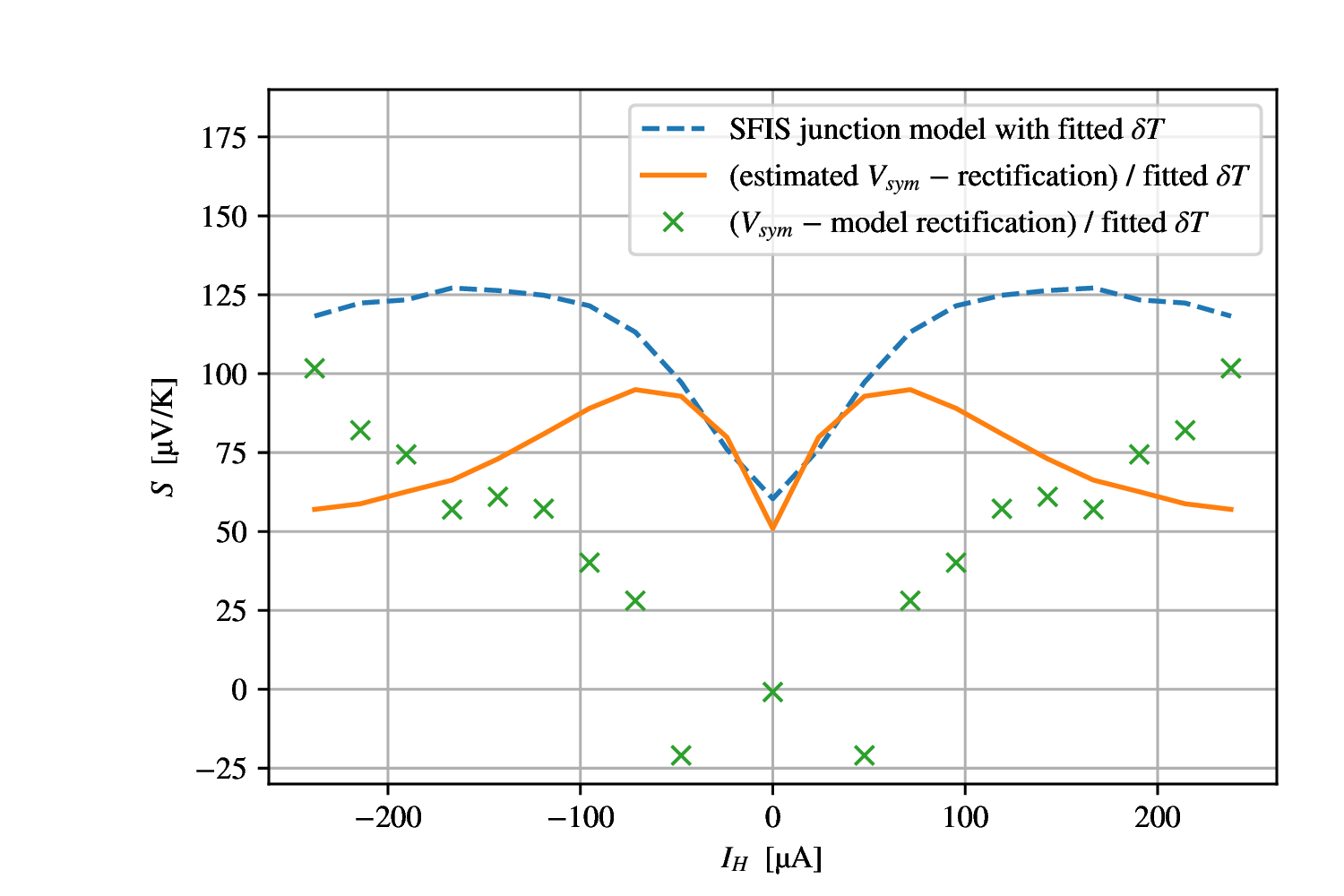}
\caption{
  \label{fig:SvsIH}
  Estimates for Seebeck coefficient.
  Obtained based on the temperatures in Fig.~\ref{fig:TNvsIH},
  and voltage from
  (i) computed from the junction model,
  (ii) dashed curve in Fig.~\ref{fig:Sthermoel},
  (iii) subtracting the rectification model from $V_{sym}$ extracted from $dI/dV$ measurements.
}
\end{figure*} 

With the parameters so obtained, we find the open-circuit voltage $V=V_N-V_S$ in the superconductor by solving
\begin{align}
    0 =
    \int_{-1/2}^{1/2}ds\, I_{NFIS}(V + s V', T_N, T_S)
\end{align}
and extract the part symmetric under inverting $V_H$.
This can be calculated with and without thermoelectric effects, i.e., determining $T_S$ from the heating model or setting $T_S=T_N$. Example of such calculations is shown in Fig.~\ref{fig:Sthermoel}. We can observe that the magnitude of thermoelectricity predicted by this model is in the range of 25\%--50\%, supporting the result obtained in an alternative way in the main text, even though the quantitative agreement is not fully complete. Moreover, we observe that the model predicts the two contributions to the symmetric voltage have opposite signs. This occurs because the heating model predicts $T_S>T_N$, since the electron-phonon coupling of Aluminum is strongly suppressed by superconductivity, and a part of the heating current tunnels in and out of the superconductor imparting direct Joule heating on it.

Based on the temperature difference obtained from the model, we show in Fig.~\ref{fig:SvsIH} the Seebeck coefficient corresponding to the voltages in Fig.~\ref{fig:Sthermoel} and temperatures in Fig.~\ref{fig:TNvsIH}. These results are all based on subtracting the counterfactual model result including only rectification, and hence the accuracy is limited to providing rough guidance of the likely order of magnitude.

Finally, we can note that the relative strength of the rectification and thermoelectricity varies depending on the junction length $L_x$. If the junction is very short, there is no transverse voltage drop or rectification, whereas if the junction is very long the rectification is large. We can estimate the length scale on which thermoelectricity starts to dominate as follows.

First, from characteristics of $I_{NFIS}$ one can observe the rectification scales with the dimensionless parameter $\sim{}V'/\Delta$ describing the transverse voltage. On the other hand, thermoelectricity scales with $\sim\delta T/T$ where $\delta T=T_N-T_S$ is the temperature difference. For thermoelectricity to be large and dominating, we then want to simultaneously have $\delta T/T\sim1$ and $eV'\ll\Delta$. In a rough estimate, under such conditions, the heat balance equation~\eqref{eq:Sheatbalance} can be approximated with
\begin{gather}
   \frac{k_B^2}{e^2 R_T} T \delta T \approx \tilde{g}\mathcal{V}_S\Sigma_S T^5
   \\
   \frac{\delta T}{T} \approx 1 \;\Rightarrow\; \tilde{g}T^3 \approx T_{x,S}^3 = \frac{k_B^2}{e^2 \rho_{\square} t_S \Sigma_S}
   \,.
\end{gather}
We assume here that the phonon system is at zero temperature. 
Here, $\tilde{g}$ is the ratio of suppression factors due to superconductivity, in the tunneling compared to that of e-ph coupling. Moreover, $R_T=\rho_\square/(L_x W)$ is the tunneling resistance where $L_x$ is the junction length, $W$ its width and $\rho_\square$ the square resistivity, and $\mathcal{V}_S=L_x W t_S$ and $\Sigma_S$ are the volume and the electron-phonon coupling in the superconductor \cite{giazotto_opportunities_2006}, and $t_S$ is the superconductor thickness. At low temperatures ($0.2$--$1.2\,$K), based on numerical calculations for $\dot{Q}_{\rm e-ph}/\dot{Q}_{\rm tun}$, we estimate $\tilde{g}\approx 2(k_BT/\Delta)^2$. 
Moreover, since $T$ is maintained above the phonon temperature by Joule heating,
\begin{gather}
   \Sigma_N T^5 \simeq \rho_N j_H^2 = \rho_N \Bigl(\frac{V'}{\rho_N L_x}\Bigr)^2
   \;\Rightarrow\;
   V' = L_x \sqrt{\rho_N \Sigma_N T^5}
   \,,
\end{gather}
where $\rho_N$, $\Sigma_N$ are the resistivity and e-ph coupling on the normal side, and $j_H$ the current density of the heating current. Finally, the condition $eV'\ll\Delta$ is equivalent to
\begin{gather}
   L_x \ll L_{x,c} 
   =   
   \sqrt{
   \frac{\Delta^2\tilde{g}/T^2}{e^2 \rho_N \Sigma_N T_{x,S}^3}
   }
   \approx
   \sqrt{\frac{2\rho_{\square} t_S \Sigma_S}{\rho_N \Sigma_N}}
   \approx{}
   100\,\mathrm{\mu{}m}
   \,.
\end{gather}
Note that the precise value depends on material parameters, also because our estimate for $\tilde{g}$ depends somewhat on values of $\Gamma$ and spin-flip scattering in the superconductor. The result is however well consistent with the fact that in the experiment of the main text, rectification dominates thermoelectricity.

In the estimates in this section, we have assumed parameter values estimated from our experiment, except for the electron-phonon coupling constants for which literature values are assumed: $\Sigma_S=2\times10^8$~W/m$^3$K$^5$, $\Sigma_N=2\times10^9$~W/m$^3$K$^5$, \cite{giazotto_opportunities_2006} $\rho_\square=0.39$~$\Omega\,$mm$^2$, $\rho_N=3.8\times10^{-8}$~$\Omega$m,$t_S=4$~nm.

%\bibliography{refs.bib}
\end{document}